# THE LYMAN-ALPHA FOREST IN THE COLD DARK MATTER MODEL

Lars Hernquist[1,2], Neal Katz[3], David H. Weinberg[4,5], Jordi Miralda-Escudé[5]

E-mail: lars@helios.ucsc.edu, nsk@astro.washington.edu, dhw@payne.mps.ohio-state.edu, jordi@sns.ias.edu

## ABSTRACT

Cosmological simulations with gas provide a detailed description of the intergalactic medium, making possible predictions of neutral hydrogen absorption in the spectra of background QSOs. We present results from a high-resolution calculation of an $\Omega = 1$ cold dark matter model. Our simulation reproduces many of the observed properties of the Ly$\alpha$ forest surprisingly well.

The distribution of HI column densities agrees with existing data to within a factor of $\sim$ two over most of the range from $10^{14}$ cm$^{-2}$ to $10^{22}$ cm$^{-2}$; i.e., from unsaturated Ly$\alpha$ forest lines to damped Ly$\alpha$ systems. The equivalent width distribution matches the observed exponential form with a characteristic width $W_* \approx 0.3$ Å. The distribution of $b$-parameters appears consistent with that derived from QSO spectra. Most of the low column density absorption arises in large, flattened structures of moderate or even relatively low overdensity, so there is no sharp distinction between the Ly$\alpha$ forest and the "Gunn-Peterson" absorption produced by the smooth intergalactic medium.

Our results demonstrate that a Ly$\alpha$ forest like that observed develops naturally in a hierarchical clustering scenario with a photoionizing background. Comparison between simulations and high-resolution QSO spectra should open a new regime for testing theories of cosmic structure formation.

*Subject headings:* quasars: absorption lines, Galaxies: formation, large-scale structure of Universe

## 1. Introduction

[1]University of California, Lick Observatory, Santa Cruz, CA 95064

[2]Sloan Fellow, Presidential Faculty Fellow

[3]University of Washington, Department of Astronomy, Seattle, WA 98195

[4]Ohio State University, Department of Astronomy, Columbus, OH 43210

[5]Institute for Advanced Study, Princeton, NJ 08540

The Ly$\alpha$ absorption systems in QSO spectra provide an excellent observational probe of the high-redshift universe. At $z \sim 2 - 5$, they easily outnumber all other detectable tracers of cosmic structure. As a tool for testing cosmological theories, Ly$\alpha$ absorbers offer several advantages: they sample a broad range of redshifts, they trace baryonic material over a wide range of densities, temperatures, and ionization states, they yield good statistical constraints on structure because of their large numbers, and they may be primitive enough to retain a more direct memory of their initial conditions than highly nonlinear objects like galaxies and quasars. Several attempts have been made to describe the formation of Ly$\alpha$ absorbers as arising within the general gravitational instability theory of cosmic structure formation (Arons 1972; Rees 1986; Ikeuchi 1986; Bond, Szalay, & Silk 1988; McGill 1990). With cosmological simulations that incorporate gravity and gas dynamics, one can make detailed predictions of Ly$\alpha$ absorption in *a priori* theoretical models (Cen et al. 1994; hereafter CMOR), while circumventing many of the idealizations required by analytic calculations.

This paper describes an analysis of the Ly$\alpha$ forest in a numerical simulation of an $\Omega = 1$ cold dark matter (CDM) model, with $H_0 = 50 \ \mathrm{km \, s^{-1} \, Mpc^{-1}}$, normalized to yield a present day rms mass fluctuation $\sigma_{16} = 0.7$ in spheres of radius 16 Mpc (close to the value advocated by White, Efstathiou & Frenk 1993). Our investigation is similar in spirit to that of CMOR, but we examine a different theoretical model ($\Omega = 1$ CDM instead of a low density model with a cosmological constant), and we use a very different numerical method, based on smoothed-particle hydrodynamics (SPH; see, *e.g.*, Lucy 1977; Gingold & Monaghan 1977; Hernquist & Katz 1989; Monaghan 1992). For this application, SPH has the advantage of providing an unusually large dynamic range in density, enabling us to study both low density, Lyman-forest systems and dense, radiatively cooled, Lyman-limit and damped Ly$\alpha$ systems in a single simulation. Resolving high densities requires short time steps, so the price of this dynamic range is a limitation in particle number, and hence in mass resolution. Our results for Lyman-limit and damped Ly$\alpha$ absorption are described in a companion paper (Katz et al. 1995, hereafter KWHM), so here we focus on systems with HI column densities below $10^{17} \mathrm{cm^{-2}}$.

## 2. Simulation

The simulation volume represents a periodic cube of comoving size 22.222 Mpc drawn randomly from a CDM universe with $\Omega = 1$ and baryon density $\Omega_b = 0.05$. The initial conditions are identical to those employed by Katz, Hernquist, & Weinberg (1992; hereafter KHW) and Hernquist, Katz, & Weinberg (1995), but the present simulation differs from our previous efforts in two important respects. First, we impose a uniform photoionizing radiation field, with intensity $J(\nu) = J_0 \left(\nu_0/\nu\right) F(z)$, where $\nu_0$ is the Lyman-limit frequency, $J_0 = 10^{-22} \, \mathrm{erg \, s^{-1} \, cm^{-2} \, sr^{-1} \, Hz^{-1}}$, and

$$F(z) = \begin{cases} 0, & \text{if } z > 6 \, ; \\ 4/(1+z), & \text{if } 3 \leq z \leq 6 \, ; \\ 1, & \text{if } 2 < z < 3 \, . \end{cases} \tag{1}$$

We compute radiative cooling and heating rates assuming optically thin gas in ionization equilibrium with this radiation field. The redshift history in equation (1) is consistent with observational constraints, but these have large uncertainties. The intensity $J_0$ is lower than most estimates; we shall comment on this point in §3.

The second important difference from the KHW simulation is a factor of eight improvement in mass resolution. We use $64^3$ SPH particles and $64^3$ dark matter particles, lowering the masses of individual particles to $1.45 \times 10^8 M_\odot$ and $2.8 \times 10^9 M_\odot$, respectively. The softening length for gravitational forces is 20 kpc in comoving coordinates (13 kpc equivalent Plummer softening). The gas resolution varies from $\sim 5$ kpc in the highest density regions to $\sim 200$ kpc in the lowest density regions. Because of the larger number of particles, we have only evolved this new simulation to $z = 2$, which is adequate for many aspects of absorption line studies. A detailed description of the simulation code, TreeSPH, and the treatment of cooling and ionization appears in Katz, Weinberg, Hernquist & Hernquist (1995). Other results from this simulation are discussed by KWHM and by Weinberg, Hernquist & Katz (1995).

## 3. Results

At each output, the simulation provides positions, velocities, densities, and temperatures of the baryonic fluid elements represented by SPH particles. It is straightforward to compute the neutral hydrogen absorption that would be produced in the light of a background QSO along an arbitrary line of sight through the simulation volume. First, we calculate the neutral fraction associated with each gas particle. We then spread each particle over its 3-dimensional SPH smoothing volume and take a line integral through the smoothed distribution to determine the neutral hydrogen mass, neutral mass-weighted velocity, and neutral mass-weighted temperature along the line of sight. Finally, we use these 1-dimensional profiles to calculate the optical depth as a function of frequency (see, e.g., the description in CMOR).

Figure 1 shows artificial spectra along four randomly chosen lines of sight at redshift $z = 2$. The physical size of the periodic simulation box is 7.41 Mpc, corresponding to a Hubble flow of 1924.5 km/s. In each panel, the solid line shows the transmission $T = e^{-\tau}$, where $\tau$ is the Ly$\alpha$ optical depth. The translation from velocity $v$ to wavelength $\lambda$ is simply $\lambda = 1216 \times (1 + z) \times (1 + v/c)$ Å, where $z = 2$.

In each panel of Figure 1, the dashed line shows a spectrum along a line of sight 100 kpc away from the "primary" spectrum represented by the solid line. The dotted line shows a spectrum at 300 kpc separation (200 kpc from the dashed spectrum). Many absorption features appear in both of the first two spectra, and there are significant matches even for lines of sight separated by 300 kpc, though these are often accompanied by substantial changes in the features' depth or shape. (For the typical absorption features in Figure 1, the SPH smoothing length of the gas particles is $h \lesssim 50$ kpc.) As shown in Figure 1 of KWHM, much of the low column density absorption arises in long, filamentary structures, explaining the correlation between separated lines of sight (see also Figure 1 of CMOR). Qualitatively, it appears that our results can account for the large coherence scale found in absorption studies of QSO pairs (e.g. Bechtold et al. 1994; Dinshaw et al. 1994, 1995), though quantitative tests (e.g. Charlton, Churchill & Linder 1995) are needed to assess the agreement or lack of agreement with recent observations.

The standard technique for identifying and measuring properties of QSO absorption lines involves fitting a spectrum with a superposition of Voigt profiles. Such a procedure is difficult to automate, and, moreover, our simulations imply that the underlying assumption that individual features are well represented as Voigt profiles is not valid in detail. We have therefore adopted a simpler prescription for measuring line properties. Moving along a transmission spectrum ($T$ vs. $\lambda$) in the direction of increasing wavelength, we identify an absorption feature when the transmission drops below some threshold $t_T$ and later rises back above it. We determine the equivalent width $W$ by integrating $(1 - T)$ between the down-crossing wavelength $\lambda_1$ and the up-crossing wavelength $\lambda_2$. We define the b-parameter to be that of a Voigt profile that would have the same equivalent width $W$ and interval width $\Delta\lambda = \lambda_2 - \lambda_1$ between the two threshold-crossing points (notice that the equivalent width here includes only the contribution between $\lambda_1$ and $\lambda_2$, and not the tails outside this interval, and it is therefore *not* the same as what is usually meant by the equivalent width of a line fitted by a Voigt profile). We separately determine the column density from the integrated optical depth between $\lambda_1$ and $\lambda_2$, using the relation $dN_{HI} = (m_e c^2)/(\pi e^2 f \lambda)\, \tau\, (d\lambda/\lambda)$, where $f = 0.416$ is the oscillator strength (Gunn & Peterson 1965). Throughout this paper we adopt a transmission threshold $t_T = 0.7$. At $z = 2$, where line-blending is not severe, our results are insensitive to the adopted threshold.

The above procedure, developed by one of us (JM), has also been applied to the simulations of Miralda-Escudé et al. (1995). We expect it to yield line properties similar to what would be obtained by Voigt-profile fitting, except at the lowest column densities, where any technique becomes sensitive to the assumptions used to separate blended features; at the same time, our column densities are physical quantities, as explained above, and may differ from the values obtained from line fits if the velocity distribution is non-gaussian. Our approach enables us to make a rough quantitative comparison between the simulation and existing absorption-line data. More precise comparisons with observational data and refinements of this procedure will be presented in future papers.

Figure 2 displays the column-density distribution $f(N) \equiv d^2N/dz\, dN_{\rm HI}$, the number of clouds per unit redshift per linear interval of HI column density. Below $10^{15.5}\,{\rm cm}^{-2}$, we obtain $f(N)$ by applying the procedure described above to 1200 random lines of sight. Above this column density we determine $f(N)$ by creating a map of the HI column density projected through the simulation cube and measuring the fractional area above each column density (KWHM). Since our box is only 7.4 Mpc deep and high column density lines are rare, the absorption along any line of sight that has $N_{\rm HI} > 10^{15.5}\,{\rm cm}^{-2}$ in the projected map is always dominated by a single absorber. Above $10^{17}\,{\rm cm}^{-2}$, it is necessary to include the effects of self-shielding when computing neutral hydrogen fractions. Our procedure for doing so, and a discussion of the physical nature of the high column density systems, appear in KWHM. Observational data and error bars in Figure 2 are taken from Table 2 of Petitjean et al. (1993).

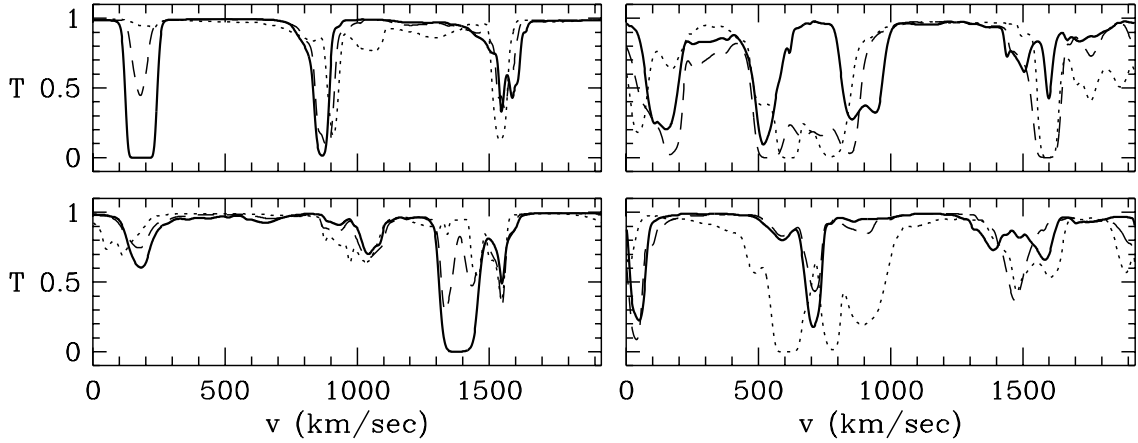

Fig. 1.— Examples of artificial spectra at $z = 2$. Solid lines show transmission against velocity along four random lines of sight. At this redshift, the physical size of the periodic simulation box is 7.41 Mpc, corresponding to a Hubble flow of 1924.5 km/s. Dashed and dotted lines show spectra along lines of sight displaced arbitrarily from that of the primary spectrum by physical separations of 100 kpc and 300 kpc, respectively.

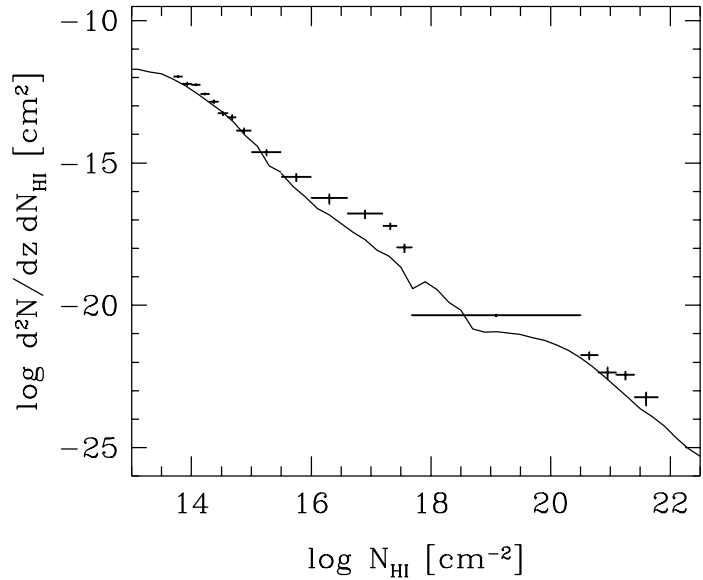

Fig. 2.— Distribution of neutral hydrogen column densities. The solid line shows the simulation results at $z = 2$. Points with error bars are taken from Petitjean et al. (1993); we multiply their values by $1 + z = 3$ to convert from number of lines per "absorption distance" interval $\Delta X$ to number of lines per redshift interval $Deltaz$.

CDM is an *a priori* theory "designed" to explain galaxy formation and large-scale structure in a universe with small microwave background anisotropies. We did not adjust any parameters to obtain the match in Figure 2. There is a significant discrepancy with the Petitjean et al. (1993) data for column densities near $10^{17} \mathrm{cm}^{-2}$, which could reflect either a failure of standard CDM or the presence in the real universe of an additional population of Lyman-limit systems that are not resolved by our simulation. However, the level of agreement over eight orders of magnitude in $N_{\mathrm{HI}}$ is still remarkable.

The results in Figure 2 depend on our choices for the mean baryon density, $\Omega_b = 0.05$, and the intensity of the ionizing background, $J_0 = 10^{-22} \mathrm{erg\, s^{-1}\, cm^{-2}\, sr^{-1}\, Hz^{-1}}$, which is lower than most estimates of the expected background from QSOs (e.g. Meiksin & Madau 1993). Changing these two quantities would change the neutral column densities (and therefore shift the distribution in Figure 2 horizontally) proportionally to $\Omega_b^2/J_0$ (neglecting any small effects from changes in the gas temperature and from the self-gravity of the baryons). Thus, if $\Omega_b = 0.1$, we would obtain a similar result to that in Figure 2 at low column densities for the more reasonable intensity $J_0 = 4 \times 10^{-22} \mathrm{erg\, s^{-1}\, cm^{-2}\, sr^{-1}\, Hz^{-1}}$, which also agrees with observations of the proximity effect (Bechtold 1994 and references therein). If future observations confirm Songaila et al.'s (1994) high value for the primordial deuterium abundance, then the implied low baryon density coupled with the minimal $J_0$ from observed QSOs may make it difficult for many cosmological models (certainly this one) to reproduce the observed level of neutral hydrogen absorption in QSO spectra, unless many of the absorbers originate not from gravitational collapse as modeled here but from some other physical mechanism, e.g. pressure-confined clouds in a shock-heated intergalactic medium (Sargent et al. 1980; Ikeuchi & Ostriker 1986).

The simulated $f(N)$ flattens below $N_{\mathrm{HI}} \sim 10^{13.8} \mathrm{cm}^{-2}$. The Petitjean et al. (1993) table does not extend below this column density, but recent analyses of high-resolution spectra from the Keck telescope suggest that $f(N)$ continues to rise as a power law down to $N_{\mathrm{HI}} \sim 2 \times 10^{12} \mathrm{cm}^{-2}$ or even lower (e.g. Songaila, Hu & Cowie 1995; Hu et al. 1995). Our identification procedure misses low column density lines, either because they are blended with more prominent absorption systems or, when they are isolated, because they fail to lower the transmission below our threshold $t_T$. An assessment of this discrepancy will therefore require an analysis of simulations and data using similar procedures. If real, the discrepancy might reflect a flaw in standard CDM, a failure of our simulation to resolve the weakest absorption systems, or the presence of absorption lines caused by other phenomena.

The distribution of equivalent widths of our lines, shown in Figure 3, is almost perfectly described by an exponential law, $n(W) \propto e^{-W/W_*}$, with characteristic width $W_* = 0.28$ Å, the same distribution as that derived by Murdoch et al. (1986) for observed lines with equivalent widths greater than 0.32 Å. For equivalent widths smaller than 0.2 Å, Murdoch et al. find a weak overabundance of lines relative to this exponential distribution. We do not find such an overabundance in the simulation; this discrepancy is related to the paucity of very low column density lines in our model, as discussed above.

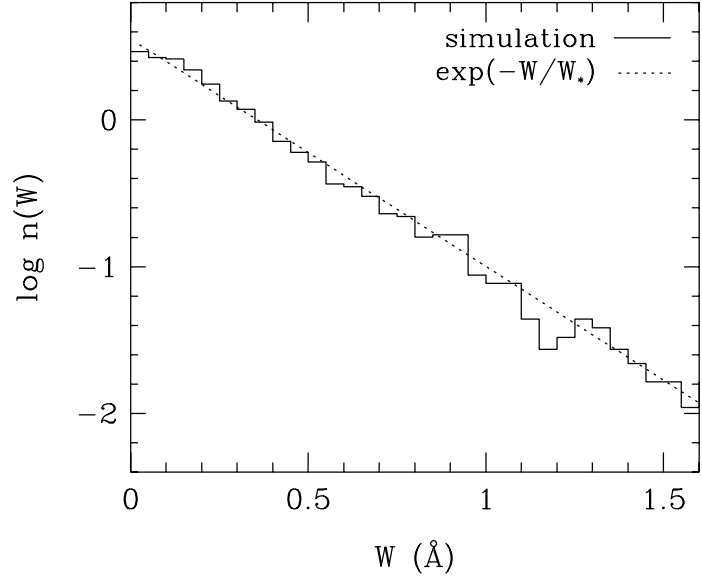

Fig. 3.— Distribution of equivalent widths of the absorption lines. Solid histogram shows the simulation result at $z = 2$. Dotted line shows an exponential distribution with characteristic width $W_* = 0.28$ Å.

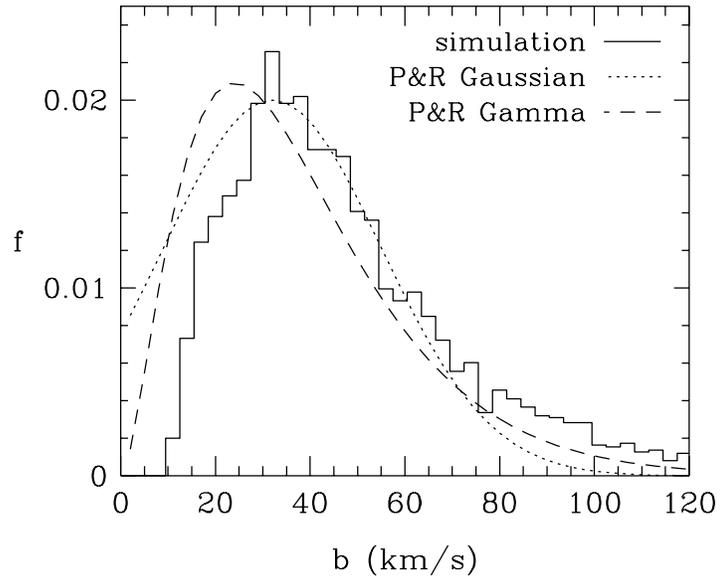

Fig. 4.— Distribution of $b$-parameters. Solid histogram shows the simulation result at $z = 2$. Dotted and dashed lines show the distributions that Press & Rybicki (1993) obtained by fitting two different functional forms to observed column density and equivalent width distributions.

As emphasized by Press & Rybicki (1993, hereafter PR), the equivalent width distribution is not controlled by the column density distribution alone, because at fixed column density the value of the $b$-parameter determines the location of the absorption line on the curve of growth. The agreement in Figures 2 and 3 thus implies that our simulated lines must have a reasonable distribution of $b$-parameters. PR proposed two different functional forms for the $b$-parameter profile and determined the free parameters of their distributions by fitting the observed $f(N)$ and $n(W)$. Figure 4 compares the histogram of $b$-parameters from our simulation analysis to the PR results. The three sets of data agree fairly well with one another, and also with the distribution obtained by Carswell (1989) from direct line-fitting.

High-resolution, high signal-to-noise spectra from the Keck telescope should improve estimates of the $b$-parameter distribution in the near future. However, this property of the lines may be particularly sensitive to the procedure used to analyze the spectra (witness the continuing controversy over the reality of lines with $b < 20$ km/s), so it will be especially important to analyze simulated and real spectra in the same manner. The $b$-parameters convey information about the physical structure of the absorbers (mainly the gas temperature and the velocity dispersion), so they may have considerable power to distinguish theoretical models.

## 4. Discussion

The agreement between the simulated and observed line populations suggests that our calculation provides a realistic general picture for the origin of the Ly$\alpha$ forest, even if the cosmological scenario and numerical realization are not correct in all their details. It appears that the Ly$\alpha$ forest can develop naturally in a hierarchical theory of structure formation with a photoionizing UV background. The high column density lines ($N_{\rm HI} > 10^{17} {\rm cm}^{-2}$) arise from radiatively cooled gas associated with forming galaxies (KWHM), in collapsed, high density regions. Low column density absorption ($N_{\rm HI} \sim 10^{13} - 10^{15} {\rm cm}^{-2}$) is produced by systems characterized by an assortment of scales and in various stages of gravitational infall and collapse. As a result, the low column density absorbers are physically diverse: they include filaments of warm gas, caustics in frequency space produced by converging velocity flows (McGill 1990), high density halos of hot, collisionally ionized gas, layers of cool gas sandwiched between shocks (CMOR), and modest local undulations in undistinguished regions of the intergalactic medium. Temperatures of the absorbing gas range from below $10^4 K$ to above $10^6 K$.

The "typical" low column density absorbers — to the extent that we can identify such a class — are flattened structures of rather low overdensity ($\rho/\bar{\rho} \sim 1 - 10$), and have $b$-parameters that are often set by peculiar motions or Hubble flow rather than thermal broadening. Gravitational confinement (Rees 1986; Ikeuchi 1986), pressure confinement by hot gas in halos (Bahcall & Spitzer 1969), and ram-pressure confinement by infalling gas (CMOR) all play significant roles, but because of their low overdensities, most of the absorbers are far from dynamical or thermal equilibrium. Many systems are still expanding with residual Hubble flow, so their physical densities and neutral fractions decrease with time. We have not quantified the evolution of the $f(N)$ histogram in this paper because our threshold algorithm tends to lose low column density lines to blending at higher redshifts. However, the mean opacity of the simulated forest climbs steadily with redshift, and the number of lines above a specified column density increases with $z$ when blending is not severe. This evolution is driven primarily by the increase in physical density with $z$, which raises the neutral fraction, and hence the opacity, of individual absorbers. The $f(N)$ distribution shifts to the right, not up.

Traditional searches for the Gunn-Peterson (1965) effect implicitly assume a uniform intergalactic medium (IGM) punctuated by discrete clouds; thus, absorption in identified lines is removed before seeking a continuum depression. Hydrodynamic simulations, on the other hand, reveal a smoothly fluctuating IGM, with no sharp distinction between "background" and "Ly$\alpha$ clouds". According to this interpretation, one might even say that the Ly$\alpha$ forest *is* the Gunn-Peterson effect, if by the latter one means absorption by neutral hydrogen in the diffuse IGM. For a generalized form of the "Gunn-Peterson test," one can abandon a distinction between lines and background and examine the full distribution function of HI optical depth, an approach we will take in a later paper. Such an analysis could reveal the signature of gas in underdense regions expanding faster than the Hubble flow (Reisenegger & Miralda-Escudé 1995).

The *mean* transmission through our simulated IGM seems in reasonable agreement with observations. Press, Rybicki & Schneider (1993), analyzing the Schneider, Schmidt & Gunn (1991) QSO spectra, find a mean Ly$\alpha$ optical depth $\bar{\tau}_\alpha = 0.0037(1+z)^{3.46}$, implying a mean transmission $\langle T \rangle = e^{-\bar{\tau}} = 0.85$, 0.64, and 0.38 at $z = 2$, 3, and 4, respectively. The simulation yields $\langle T \rangle = 0.83$, 0.63, and 0.31 at the same redshifts. The $\sim 3\sigma$ discrepancy at $z = 4$ would be eliminated if we kept the UV background intensity constant between $z = 4$ and $z = 3$ instead of following equation (1).

CMOR and Miralda-Escudé et al. (1995) find that a flat, low density CDM model ($\Omega = 0.4$, $\Omega_\Lambda = 0.6$, $H_0 = 65$ km s$^{-1}$ Mpc$^{-1}$, $\Omega_b = 0.0355$, COBE-normalized) can also reproduce the essential properties of the Ly$\alpha$ forest, including the abundance of low column density systems, with $\Omega_b = 0.036$ and the same background intensity $J_0 = 10^{-22}$ erg s$^{-1}$ cm$^{-2}$ sr$^{-1}$ Hz$^{-1}$ as employed here. Zhang, Anninos & Norman (1995) achieve similar success with a COBE-normalized, $\Omega = 1$ CDM model, using a higher background intensity. (For related studies using approximate treatments of hydrodynamics see Petitjean, Mücket & Kates 1995, Mücket et al. 1995, and Bi, Ge, & Fang 1995.) Clearly, much further work is needed to determine which cosmological models can match the observed Ly$\alpha$ forest and which cannot. The success of three somewhat different models in reproducing $f(N)$ suggests that the column density distribution alone may be a rather blunt test, serving largely to constrain the parameter combination $J_0/\Omega_b^2$ within a specific theory. However, one can check to see whether the required evolution of $J_0$ is physically reasonable, and once this parameter combination is fixed, there is not much freedom to adjust the predicted higher-order properties of the Ly$\alpha$ forest — $b$-parameters, line shapes, clustering, transmission distributions, Ly$\beta$ absorption, and so forth. We can, therefore, expect that QSO absorption lines will assume an important role in testing theories of cosmic structure formation.

We acknowledge helpful discussions with Jill Bechtold, Renyue Cen, Craig Hogan, Jeremiah Ostriker, Max Pettini, Tom Quinn, Martin Rees, Wal Sargent, David Tytler, and Ray Weymann. This work was supported in part by the Pittsburgh Supercomputing Center, the National Center for Supercomputing Applications (Illinois), the San Diego Supercomputing Center, the Alfred P. Sloan Foundation, NASA Theory Grants NAGW-2422, NAGW-2523, and NAG5-2882, NASA HPCC/ESS Grant NAG 5-2213, NASA grant NAG-51618, and the NSF under Grant ASC 93-18185 and the Presidential Faculty Fellows Program. DHW acknowledges the support of a Keck fellowship at the Institute for Advanced Study during early phases of this work, and JM acknowledges support from the W. M. Keck Foundation.